\documentclass[aps,prd,reprint,groupaddress,noshowpacs,noshowkeys]{revtex4-1}

\usepackage{amsmath,amssymb,mathrsfs}

\usepackage{graphicx}% Include figure files
\usepackage{dcolumn}% Align table columns on decimal point
\usepackage{bm}% bold math
\usepackage{xcolor}
\usepackage[T1]{fontenc}
\usepackage[pdftex,
  hidelinks,
  colorlinks=true,
  linkcolor=black,
  anchorcolor=black,
  citecolor=black,
  urlcolor=black,
  pdfauthor= {Michael Glinsky},
  pdfsubject = {Helicity in Hamiltonian dynamical systems},
  pdfdisplaydoctitle,
  bookmarks=true,
  bookmarksopen=true]{hyperref}

\begin{document}
\title{Helicity in Hamiltonian dynamical systems}

\author{Michael E. Glinsky}
\affiliation{BNZ Energy Inc., Santa Fe, NM, USA}

\author{Poul G. Hjorth}
\affiliation{DTH, Denmark}

%\date{\today, DRAFT, version 1}

\begin{abstract}
Helicity plays a unique role as an integral invariant of a dynamical system.  In this paper, the concept of helicity in the general setting of Hamiltonian dynamics is discussed.   It is shown, through examples, how the conservation of overall helicity can imply a bound on other quantities of the motion in a nontrivial way.
\end{abstract}

\maketitle

\onecolumngrid

\section{Introduction}\label{sec:intro}
The total helicity of a vector field has been studied in the past by various authors \citep{carter1980active,berger1984topological,moffatt1969degree,taylor1986relaxation,arnold1974asymptotic,freedman1988note}. The flows (diffeomorphisms in general) for which helicity has been studied have usually been volume preserving.  Under such flows, the total helicity of a vector field is conserved provided that the manifold on which the vector field is defined is closed or a boundary condition is met \citep{berger1984topological}. This has been used to put a lower bound on the enstrophy \citep{moffatt1969degree}.  A practical application of this has been to determine the equilibrium magnetic field profile in a reversed field pinch fusion device \citep{taylor1986relaxation}.

A more advanced treatment has been done by others relating helicity to a quadratic asymptotic linking number of the field lines associated with the vector field \citep{arnold1974asymptotic}. More recently constraints imposed by a general topological linkage of field lines has been investigated \citep{freedman1988note}.

In this paper, we examine helicity in terms of the Poincar\'e one and two forms for a single particle.  The state of a continuum system is represented by a graph in the phase space for a single particle \citep{spiegel1982mathematical}.  This graph defines a three dimensional manifold.  The helicity of the system is defined as the integral of a three form over the graph.  The evolution of the state of the system (\textit{i.e.,} graph) is described by a flow on the phase space.  If the flow is Hamiltonian and certain boundary conditions on the graph are met, helicity is found to be conserved.  This allows us to apply helicity conservation to any Hamiltonian dynamical system, including those with a Hamiltonian which is time dependent.  Under such flows on compact manifolds, with or without boundary, we find that helicity conservation puts bounds on the related quantities -- enstrophy and solenoidal energy.

In Sec.~\ref{sec:defs}, helicity is defined and the conditions under which it is conserved are discussed.  The pullback of helicity into a familiar vector expression appears in Sec.~\ref{sec:pullback}.  The manner in which helicity puts bounds on enstrophy and solenoidal energy is shown in Sec.~\ref{sec:con_var}.  This bound is determined by the smallest eigenvalue of the curl operator.  Section~\ref{sec:eigen} presents solutions to the eigenvalue problem for two different examples, one without boundary and another with boundary.  Shown in Sec.~\ref{sec:phys_example} are various examples of how the dynamics of a continuum system, which has an infinite number of degrees of freedom, can be related to the evolution of a graph on the phase space of a single particle.  These include a perfect fluid and magnetohydrodynamics (MHD).  It is also shown how cross helicity \citep{woltjer1958theorem}, a concept related to helicity, is conserved under MHD.

\section{Basic definitions and helicity conservation}\label{sec:defs}
We begin by showing how helicity may be defined in a Hamiltonian system.  Consider a dynamical system with an $n$ dimensional configuration manifold $M^n$, phase space $T^*M^n$ and the Poincar\'e nondegenerate 2-form $\Omega^2$ defined on the phase space \citep{arnold1978mathematical}.  Let the dynamical system be described by the Hamilton function $H=H(p,q)$.  The associated equation of motion is
\begin{equation}
\label{eqn:ham_equation}
i_\mathbf{u} \Omega^2 = -dH.
\end{equation}
Here,
\begin{equation}
\Omega^2 \equiv dp \wedge dq = d \Lambda^1
\end{equation}
is the canonical version of the nondegenerate 2-form, where $\Lambda = p \, d q$ is the Poincar\'e 1-form, and
\begin{equation}
\mathbf{u} = \left[ \begin{array}{c} \dot{p} \\ \dot{q} \end{array} \right] = \frac{dp}{dt} \frac{\partial}{\partial p} + \frac{dq}{dt} \frac{\partial}{\partial q} = u^p \frac{\partial}{\partial p} + u^q \frac{\partial}{\partial q}
\end{equation}
is the tangent vector to the flow in phase space.  From the Cartan formula (valid for any $n$-form $\omega$)
\begin{equation}
\mathscr{L}_\mathbf{u}\omega =d(i_\mathbf{u} \omega) + i_\mathbf{u} (d \omega),
\end{equation}
we get that
\begin{equation}
\label{eqn:lie_zero}
\mathscr{L}_\mathbf{u}\Omega^2 =d(i_\mathbf{u} \Omega^2) + i_\mathbf{u} (d \Omega^2) = 0
\end{equation}
using Eq.~\ref{eqn:ham_equation} and the fact that $d \Omega^2 = 0$.  This is the fundamental expression that Hamiltonian flow conserves phase space area (``Liouville's Theorem'').  In fact, $\Omega^2$ is only the first of a sequence of invariants ($\Omega^2$, $\Omega^2 \wedge \Omega^2$, $\dots$, $\Omega^{2n}$), where $n$ is the dimension of the configuration space.  That each of these are invariant follows from Eq.~\ref{eqn:lie_zero} and the distributive law for the Lie derivative.

We now define the helicity 3-form $K^3$ as
\begin{equation}
K^3 \equiv \Lambda^1 \wedge \Omega^2.
\end{equation}
Consider now a finite 3-volume $V$ in phase space.  The total amount of helicity contained in $V$ is given by
\begin{equation}
\mathscr{H} \equiv \int_V{K^3}.
\end{equation}
The rate of change of $\mathscr{H}$ as $V$ is evolved forward with the flow is given by
\begin{equation}
\begin{split}
\frac{d \mathscr{H}}{d t} &= \int_V{\mathscr{L}_\mathbf{u} K^3} = \int_V{(\mathscr{L}_\mathbf{u} \Lambda^1) \wedge \Omega^2 + \Lambda^1 \wedge (\mathscr{L}_\mathbf{u} \Omega^2)} \\
&= \int_V{\left( i_\mathbf{u} \Omega^2 + d(i_\mathbf{u} \Lambda^1) \right) \wedge \Omega^2} = \int_V{d(i_\mathbf{u} \Lambda^1 - H) \wedge \Omega^2} \\
&= \int_V{d \left[ (i_\mathbf{u} \Lambda^1 -H) \Omega^2 \right]} = \int_{\partial V}{(i_\mathbf{u} \Lambda^1 -H) \Omega^2} = \int_{\partial V}{L \, \Omega^2}.
\end{split}
\end{equation}
Here, one can declare $L$ to be the Lagrange function.  From this equation, we can conclude several things.  If $L \Omega^2$ vanishes on the boundary $\partial V$ of the volume $V$ or if $V$ has no boundary, then the total amount of helicity inside $V$ remains constant as $V$ is carried along with the phase flow.  (Note that the condition that $\Omega^2$ vanish on the boundary of $V$ is only an initial condition since $\mathscr{L}_\mathbf{u} \Omega^2 = 0$.)  Moreover, $K^3$ is only the first of a sequence of generalized helicities $(K^3,K^5,\dots,K^{2n-1})$. For each of these, say $K^{2j+1}$, the above conservation argument holds when $\Omega^2$ is replaced by $\Omega^{2j}$ and the 3-volume is replaced by a ($2j$+1)-volume.

\section{Pullback}\label{sec:pullback}
\begin{figure}[ht]
\center\includegraphics[width=20pc]{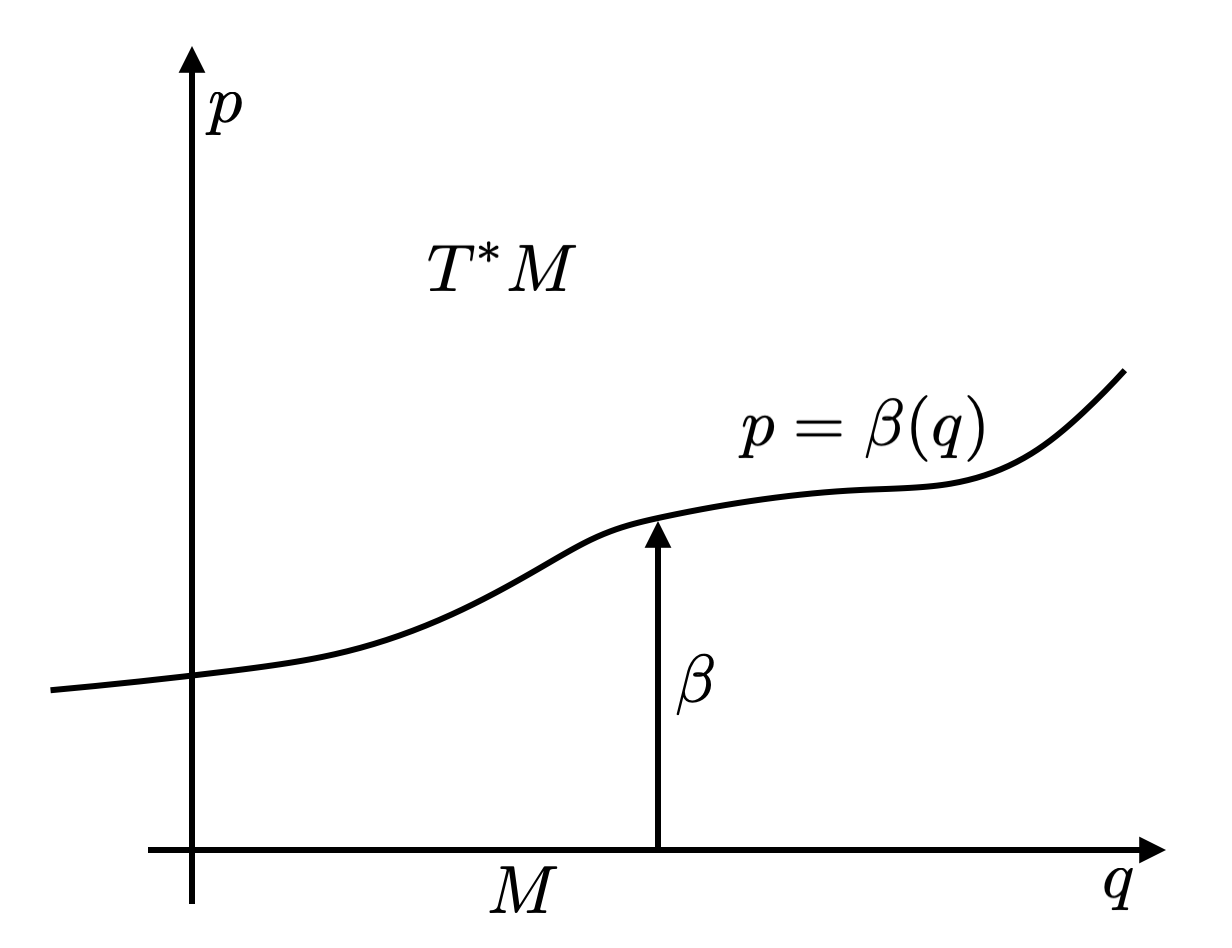}
\caption{\label{fig:pullback}Graph $V$ in phase space $T^*M$.}
\end{figure}

Locally, we can always specify an integral curve of $\mathbf{u}$ by the map from the configuration manifold $M$ into its phase space $T^*M$ which assigns to a point $q$ the corresponding momentum $p=\beta(q)$.  Such a map is shown in Fig.~\ref{fig:pullback}.  To see what helicity conservation looks like inside the configuration manifold, we pullback the forms from $T^*M$ down to $M$ with $\beta^*$.  Since $\beta^*$ respects both the exterior differential operator ``$d$'' and the wedge product ``$\wedge$'', we simply have that $\beta^* \Lambda^1 = p(q) \text{dq}$ and that, consequently, $\beta^* \Omega^2 = d(\beta^* \Lambda^1) = dp(q) \wedge dq$.  For the special case where $V$ is inside a 4-dimensional configuration manifold with coordinates $q=(t,x,y,z)$;  we get, in particular,
\begin{equation}
\begin{split}
\beta^* \Omega^2 =& \left( \frac{\partial p_z}{\partial y} - \frac{\partial p_y}{\partial z} \right) dy \wedge dz +  \left( \frac{\partial p_x}{\partial z} - \frac{\partial p_z}{\partial x} \right) dz \wedge dx + \left( \frac{\partial p_y}{\partial x} - \frac{\partial p_x}{\partial y} \right) dx \wedge dy \\
&+ \left[ \left( \frac{\partial p_0}{\partial x} - \frac{\partial p_x}{\partial t} \right) dx + \left( \frac{\partial p_0}{\partial y} - \frac{\partial p_y}{\partial t} \right) dy  +\left( \frac{\partial p_0}{\partial z} - \frac{\partial p_z}{\partial t} \right) dz  \right] \wedge dt
\end{split}
\end{equation}
or get, with notation from 3-space vector analysis,
\begin{equation}
\beta^* \Omega^2 = \omega^2_{\; \nabla \times \mathbf{p}} + \left( \omega^1_{\; \nabla p_0 - \partial_t \mathbf{p}} \right) \wedge dt,
\end{equation}
where $\omega^2_\mathbf{a} = i_\mathbf{a} \text{vol}^3$ and $\omega^1_\mathbf{a} = *(i_\mathbf{a} \text{vol}^3 )$ are the standard representation of vectors as one and two-forms.  The ``$*$'' in the definition of the 1-form $\omega^1$ is the Hodge star operator \citep{von2009differential}.  Consequently,
\begin{equation}
\begin{split}
\beta^* K^3 =& \beta^* \Lambda^1 \wedge \beta^* \Omega^2 \\
=& \left[ p_x \left(\partial_y p_z - \partial_z p_y \right) + p_y \left(\partial_z p_x - \partial_x p_z \right) + p_z \left(\partial_x p_yy - \partial_y p_x \right) \right] \text{vol}^3 \\
&+ p_0 \left[ \left(\partial_y p_z - \partial_z p_y \right) dy \wedge dz + \left(\partial_z p_x - \partial_x p_z \right) dz \wedge dx +\left(\partial_x p_y - \partial_y p_x \right) dx \wedge dy \right] \wedge dt \\
&+ \left[ \left(p_y \partial_z p_0 - p_z \partial_y p_0 \right) dy \wedge dz + \left(p_z \partial_x p_0 - p_x \partial_z p_0 \right) dz \wedge dx +\left(p_x \partial_y p_0 - p_y \partial_x p_0 \right) dx \wedge dy \right] \wedge dt  \\
&- \left[ \left(p_y \partial_t p_z - p_z \partial_t p_y \right) dy \wedge dz + \left(p_z \partial_t p_x - p_x \partial_t p_z \right) dz \wedge dx +\left(p_x \partial_t p_y - p_y \partial_t p_x \right) dx \wedge dy \right] \wedge dt \\
=& \left( \mathbf{p} \cdot \vec{\nabla} \times \mathbf{p} \right) \text{vol}^3 + \left( \omega^2_{\; p_0 \vec{\nabla} \times \mathbf{p} + \mathbf{p} \times \vec{\nabla} p_0 - \mathbf{p} \times \partial_t \mathbf{p}} \right) \wedge dt.
\end{split}
\end{equation}
In particular, we see that inside 3-space helicity density is the scalar product $\mathbf{p} \cdot \vec{\nabla} \times \mathbf{p}$.  We wish to emphasize that $\mathbf{p}$ denotes the canonical momentum.  It may be relativistic and have a magnetic component.  For the case of a perfect fluid, $\mathbf{p} = \mathbf{v}$ and helicity density is $\mathbf{v} \cdot \vec{\nabla} \times \mathbf{v}$.  In MHD, helicity density is $\mathbf{A} \cdot \mathbf{B}$, formed from the vector potential $\mathbf{A} $ and the magnetic field $\mathbf{B}$.  We will discuss this further in Sec.~\ref{sec:phys_example}.

\section{Constrained variations}\label{sec:con_var}
In this section, we discuss how the conservation of helicity, as viewed from 3-space, can set bounds on other global quantities of the motion.  In the calculations below, we suppress the pullback $\beta^*$ operator which is understood to act on all forms.  Consider the total \textit{enstrophy}, defined as the volume integral over enstrophy density:
\begin{equation}
\mathscr{N} \equiv \int_V{\Omega \wedge *\Omega}.
\end{equation}
To find the extremal value of this quantity, under the constraint of conserved helicity,  we perform a variational calculation with a real parameter $\mu$ as a Lagrange multiplier:
\begin{equation}
0 = \delta \int_V {\Omega \wedge *\Omega - \mu \, \Lambda \wedge \Omega}.
\end{equation}
Using that $\Omega = d \Lambda$ and the Leibnitz rule $d(a^p \wedge b^q) = da^p \wedge b^q + (-)^p a^p \wedge db^q$, we get that
\begin{equation}
\begin{split}
0 &= \int_V {2 \, \delta \Omega \wedge *\Omega - \mu \left( \delta \Lambda \wedge \Omega + \Lambda \wedge \delta \Omega \right)} \\
&= \int_V {2 \, \delta \Lambda \wedge \left( d *\Omega - \mu \Omega \right) + d \left[ \delta \Lambda \wedge \left(2 *\Omega - \mu \Lambda \right) \right]}.
\end{split}
\end{equation}
We now assume that $V$ is closed or that the second term $\delta \Lambda \wedge \left(2 *\Omega - \mu \Lambda \right)$ vanishes on the boundary; the variational condition then gives that enstrophy is extremized when
\begin{equation}
\label{eqn:eigen_curl}
0 = d *\Omega - \mu \Omega,
\end{equation}
\textit{i.e.,} the vector associated with the 2-form $\Omega$ must be an eigenvector of the curl operator with eigenvalue $\mu$.

The meaning of the eigenvalue or Lagrange multiplier $\mu$ can be seen by evaluating the enstrophy when the eigenvalue equation is satisfied:
\begin{equation}
\mathscr{N} = \int_V {\Omega \wedge *\Omega} = \int_V {\Omega \wedge \left( \mu d^{-1} \Omega \right)} = \mu \int_V {\Omega \wedge \Lambda} = \mu \mathscr{H}.
\end{equation}
It is the ration $\mu = \mathscr{N} / \mathscr{H}$.

One can also consider the total energy,
\begin{equation}
\mathscr{T} \equiv \int_V {\Lambda \wedge *\Lambda},
\end{equation}
and its extremal value under the constraint of conserved helicity by a similar variational calculation:
\begin{equation}
\begin{split}
0 &= \delta \int_V {\Lambda \wedge *\Lambda - \frac{1}{\mu} \Lambda \wedge \Omega} \\
&= \int_V {2 \, \delta \Lambda \wedge \left( *\Lambda - \frac{1}{\mu} d \Lambda \right) + d \left( \frac{1}{\mu} \Lambda \wedge \delta \Lambda \right)}.
\end{split}
\end{equation}
We assume that $V$ is closed or $\Lambda \wedge \delta \Lambda$ vanishes on the boundary;  the variational calculation then gives that energy is extremized when
\begin{equation}
0 = *\Lambda - \frac{1}{\mu} d \Lambda,
\end{equation}
\textit{i.e.,} the vector associated with the 2-form $*\Lambda$ must be an eigenvector of the curl operator.

The meaning of this eigenvalue $\mu$ can be seen by evaluating the energy when the eigenvalue equation is satisfied:
\begin{equation}
\mathscr{T} = \int_V {\Lambda \wedge *\Lambda} = \int_V {\Lambda \wedge \left( \frac{1}{\mu} \Omega \right)} = \frac{1}{\mu} \mathscr{H}.
\end{equation}
It is the ration $\mu = \mathscr{H} / \mathscr{T}$.

The problem of extremization of both $\mathscr{T}$ and $\mathscr{N}$ under the constraint of constant $\mathscr{H}$ has now been reduced to finding eigen 2-forms and eigenvalues of the operator $d*$.  To do this formally, let us consider the Hilbert space of all $p$-forms on a Riemannian $M^n$, $\Lambda^p(M^n)$, with the inner product
\begin{equation}
\left( \alpha^p , \beta^p \right) = \int_M {\alpha^p \wedge *\beta^p}.
\end{equation}
If $M^n$ is compact or $\alpha$ and $\beta$ have compact support, the adjoint of $d$ can be defined to be
\begin{equation}
d^A \equiv - (-)^{n(p+1)} *d*.
\end{equation}

The Hilbert space $\Lambda^p(M^n)$ admits the Hodge decomposition \citep{warner2013foundations}
\begin{equation}
\Lambda^p(M^n) = d \Lambda^{p-1}(M^n) \oplus d^A \Lambda^{p+1}(M^n) \oplus \mathcal{H}_p(M^n),
\end{equation}
where $\mathcal{H}_p(M^n)$ is the space of all harmonic $p$-forms on $M^n$, $d \Lambda^{p-1}(M^n)$ is the space of all exact $p$-forms on $M^n$, and $d^A \Lambda^{p+1}(M^n)$ is the space of all co-exact p-forms on $M^n$.  If one restricts the domain of $d^A$ to $d \Lambda^{p-1}(M^n)$, it can be easily shown by use of the ``\textit{closed-graph} theorem'' \citep{taylor1958introduction,reed1980methods} that a unique $(d^A)^{-1}$ exists and is bounded so that $\exists \; m>0$ for which
\begin{equation}
m \left\| \alpha^p \right\| \le \left\| d^A \alpha^p \right\|, \qquad \forall \alpha^p \in d \Lambda^{p-1}(M^n).
\end{equation}
Furthermore since $\left\| \alpha^p \right\| = \left\| * \alpha^p \right\|$ and $d* = \pm *d^A$,
\begin{equation}
m \left\| \alpha^p \right\| \le \left\| d * \alpha^p \right\|, \qquad \forall \alpha^p \in d \Lambda^{p-1}(M^n).
\end{equation}

Let us now consider the curl operator restricted to such a domain:
\begin{equation}
d*: \; d \Lambda^1(M^3) \to d \Lambda^1(M^3).
\end{equation}
The eigen 2-forms of $d*$ restricted to $d \Lambda^1(M^3)$ are $\zeta^2$ such that
\begin{equation}
\zeta^2 \in d \Lambda^1(M^3) \quad \text{and} \quad d*\zeta^2 = \mu \zeta^2.
\end{equation}
It is now evident from the above remarks that $m \le |\mu|$.  This conclusion that the spectrum of $d*$ on $d \Lambda^1(M^3)$ is bounded away from zero (\textit{i.e.,} $(d*)^{-1}$ exists and is compact) can now be applied to the constrained optimization of $\mathscr{T}$ and $\mathscr{N}$.

Since by hypothesis $\Omega = d \Lambda \in d \Lambda^1(M^3)$, one can immediately conclude for the eigen 2-forms of Eq.~\ref{eqn:eigen_curl} that $|\mu| \ge m$ or $\mathscr{N} \ge m |\mathscr{H}|$.  Enstrophy has a lower bound.  It is slightly more complicated for the case of $\mathscr{T}$.  Let us decompose $*\Lambda$ into $*\Lambda = d\alpha^1 + d^A\beta^3 + h^2$, where $d\alpha^1$ is the \textit{solenoidal} or \textit{transverse} component, $d^A \beta^3$ is the \textit{irrotational} or \textit{longitudinal} component, and $h^2$ is the \textit{harmonic} part.  Define the solenoidal energy and helicity to be
\begin{equation}
\mathscr{T}_0 \equiv \int_V {*d \alpha \wedge d \alpha}
\end{equation}
and
\begin{equation}
\mathscr{H}_0 \equiv \int_V { *d \alpha \wedge d*d \alpha}.
\end{equation}
Since $d \alpha^1 \in d \Lambda^1(M^3)$, $\mathscr{T}_0 \le (1/m) \, |\mathscr{H}_0|$.  Furthermore, the solenoidal part of $\Lambda$ is the only component of $\Lambda$ contribute to the helicity, therefore $\mathscr{H} = \mathscr{H}_0$ and $\mathscr{T}_0 \le (1/m) \, |\mathscr{H}|$.  The conclusion that can be drawn is that while the energy $\mathscr{T}$ has no bound, the \textit{solenoidal} energy $\mathscr{T}_0$ is bound from above under the constraint of helicity conservation.

One must be careful when applying the bounds on enstrophy and solenoidal energy.  While helicity is conserved regardless of whether $M^n$ is Riemannian or not, the bound on the eigenvalues of $d*$ can only be proved if $M^n$ is Riemannian.  This precludes using the eigenvalue bound for situations which do not have a Riemannian metric (\textit{e.g.,} Minkowski space), unless $M^n$ is always restricted to a submanifold which has a Riemannian metric (\textit{e.g.,} a volume in Minkowski space with constant time).

Note what happens to the eigenvalue bound if we extend or restrict the configuration manifold.  Suppose the new manifold $M^3$ is a subset of the old manifold $M_0^3$, that is, $M^3 \subset M_0^3$.  For every form $\alpha$ on $M^3$, a form $\alpha_0$ on $M^3_0$ can be chosen so that $\alpha_0$ equals $\alpha$ when restricted to $M^3$.  The set of forms $\alpha$ on $M^3$ such that $d*\alpha=\mu \alpha$ is contained in the set of $\alpha_0$ on $M^3_0$ with $d*\alpha_0 = \mu \alpha_0$.  Consequently, the eigenvalue bound $m$ on $M^3$ will be larger than the eigenvalue bound $m_0$ on $M_0^3$, that is,
\begin{equation}
m \ge m_0.
\end{equation}
A bound on $\mathscr{N}$ and $\mathscr{T}_0$ associated with flows which change $M^3$ can now be found if $M^3$ is always a subset of some manifold $M_0^3$.  This is the case for systems required to remain in a ``box''.

A technical detail we now address is how to extend $\alpha$ into $\alpha_0$ on a closed $M_0^3$.  Take the $C^\infty$ extension $\alpha_{\text{ext}}$ of $\alpha$ into $N^3$, a surrounding neighborhood of $M^3$ contained in $M_0^3$.  Multiply this extension by a $C^\infty$ function $f$ which is 1 on $M^3$ and is equal to zero on $M_0^3 - N^3 - M^3$.  Now, let
\begin{equation}
\beta_0 = \left\{ \begin{array}{ll}  \alpha & \textrm{ on $M^3$} \\ f \alpha_\text{ext} & \textrm{on $N^3$} \\ 0 & \textrm{on $M^3_0 - N^3 - M^3$} \end{array} \right. .
\end{equation}
The form $\beta_0$ can be decomposed into $\beta_0 = \alpha_0 +\gamma_0 + h_0$, where $\alpha_0$ is exact, $\gamma_0$ is co-exact, and $h_0$ is harmonic.  The exact component of $\beta_0$, $\alpha_0$, is equal to $\alpha$ on $M^3$ and, by definition, is an element of $d \Lambda^1(M_0^3)$.  This is exactly what we wished to find.

\begin{figure}[ht]
\center\includegraphics[width=20pc]{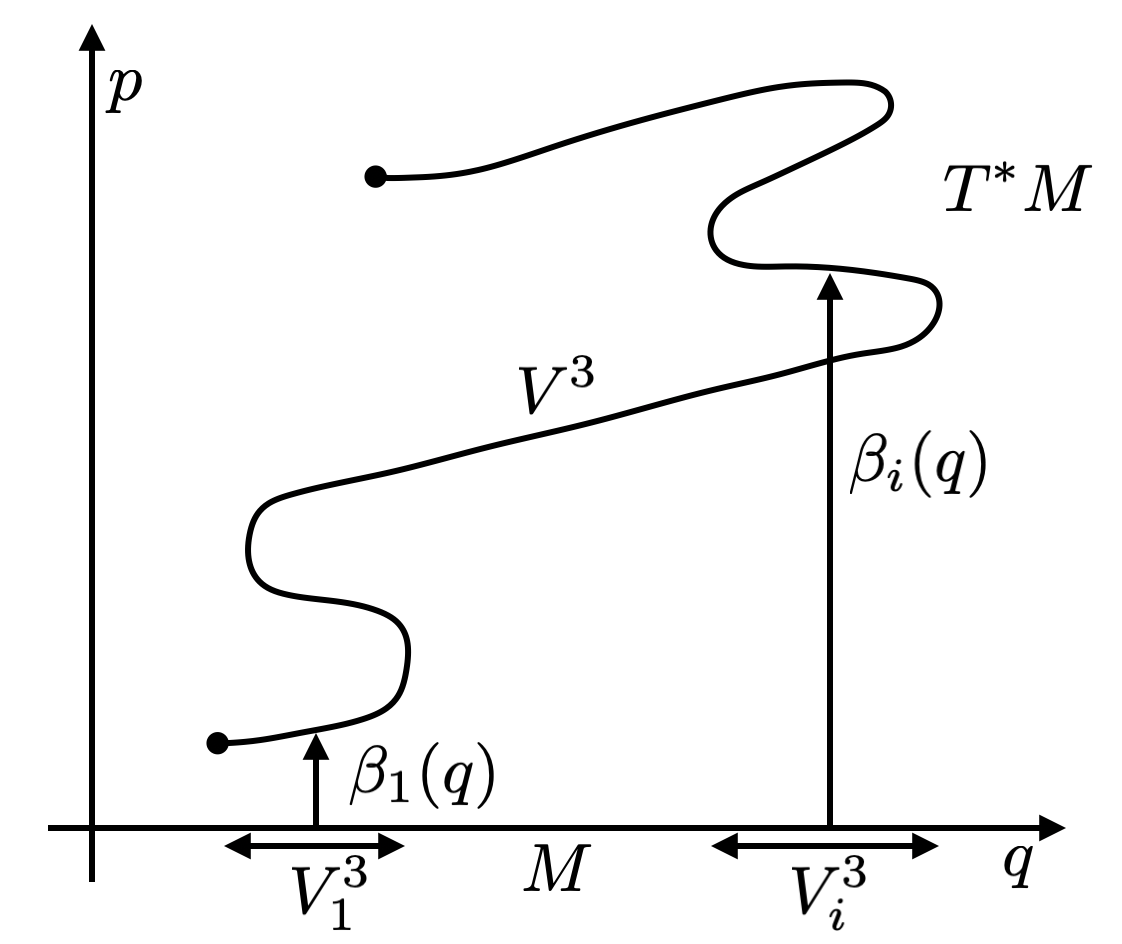}
\caption{\label{fig:multivalued}Multivalued graph, $V$.}
\end{figure}

One last technical detail is the fact that $\beta(q)$ discussed in Sec.~\ref{sec:pullback} is not always single valued.  In this case, shown in Fig.~\ref{fig:multivalued}, separate $\beta_i$ can be defined that are all single valued.  Each $\beta_i$ is defined on the domain $V_i^3$.  One can now define integration over $V^3$ as
\begin{equation}
\int_{V^3} {\alpha^3} = \sum_i {\int_{V_i^3} {\beta_i^* \alpha^3}}.
\end{equation}
Repeating the variational calculation of Sec.~\ref{sec:con_var}, one obtains the following equation for the extremal 2-forms
\begin{equation}
d * \alpha_i^2 = \mu_i * \alpha_i^2 \qquad \textrm{$\forall \alpha_i^2$ on $V_i^3$}.
\end{equation}
If the dynamics are constrained such that $V_i^3 \subset M_0^3 \; \forall i$, then by the above extension theorem $|\mu_i| > m_0 \; \forall \mu_i$.  The bound on enstrophy changes to
\begin{equation}
\mathscr{N} = \sum_i {\mu_i \mathscr{H}_i} \ge m_0 \sum_i {|\mathscr{H}_i|} \ge m_0 |\mathscr{H}|,
\end{equation}
where $\mathscr{H}_i$ is the contribution to the helicity from $\beta_i$ on $V_i^3$.  By a similar argument,
\begin{equation}
\mathscr{T}_0 \le \frac{1}{m_0} |\mathscr{H}|.
\end{equation}

\section{Eigenvalue problem}\label{sec:eigen}
The existence of bounds on $\mathscr{N}$ and $\mathscr{T}_0$ have a simple heuristic explanation.  Let the dominant (initial) contribution to all quantities be at a length scale $L_i$.  Now change this dominant length scale to the (final) value $L_f$.  By dimensional analysis, it is evident that
\begin{equation}
\frac{\mathscr{T}_f / \mathscr{T}_i}{\mathscr{H}_f / \mathscr{H}_i} \sim \frac{L_f}{L_i} \quad \text{and} \quad \frac{\mathscr{N}_f / \mathscr{N}_i}{\mathscr{H}_f / \mathscr{H}_i} \sim \frac{L_i}{L_f}.
\end{equation}
If $\mathscr{H}_f = \mathscr{H}_i$, it is obvious that one would like to let $L_f \to \infty$ to maximize $\mathscr{T}_0$ and minimize $\mathscr{N}$.  Since the size of configuration space is limited in practice to be no larger than some $L_0$, an upper bound is put on $\mathscr{T}_0$ and a lower bound is put on $\mathscr{N}$.

Let us now look at a simple example of a physical system which displays the behavior we have just discussed.  Consider the manifold $B^3$ which is a box with periodic boundary conditions and sides of length $L_x$, $L_y$ and $L_z$.  This is a closed manifold with the topology of the three torus:  $B^3 = S^1 \times S^1 \times S^1$.

Decompose an arbitrary vector field on $B^3$ into its Fourier components
\begin{equation}
\mathbf{v}(\mathbf{q}) = \text{Re} \left( \sum_{\mathbf{k} \alpha} {A_{\mathbf{k} \alpha} \mathbf{a}_{\mathbf{k} \alpha} \text{e}^{i \mathbf{q} \cdot \mathbf{k}}} \right)
\end{equation}
where
\begin{equation}
\mathbf{k} = 2 \pi \left(\frac{n_x}{L_x},\frac{n_y}{L_y},\frac{n_z}{L_z} \right)
\end{equation}
and $n_x, n_y, n_z = 0,1,2,\dots$.  The vectors $\mathbf{a}_{\mathbf{k} \alpha}$ are complex valued with unit norm and the constants $A_{\mathbf{k} \alpha}$ are complex numbers.  We wish to find $\mathbf{a}_{\mathbf{k} \alpha}$ such that $\mathbf{p}_{\mathbf{k} \alpha} = A_{\mathbf{k} \alpha} \mathbf{a}_{\mathbf{k} \alpha} \text{e}^{i \mathbf{q} \cdot \mathbf{k}}$ are eigenvectors of the curl operator, that is,
\begin{equation}
\vec{\nabla} \times \mathbf{p}_{\mathbf{k} \alpha} = \mu_{\mathbf{k} \alpha}  \mathbf{p}_{\mathbf{k} \alpha}.
\end{equation}
This can be reduced to the matrix eigenvalue equation
\begin{equation}
\left(\overleftrightarrow{G}_{\hat{\mathbf{k}}} - \lambda_{\hat{\mathbf{k}} \alpha} \overleftrightarrow{I} \right)  \mathbf{a}_{\hat{\mathbf{k}} \alpha} = 0
\end{equation}
where
\begin{equation}
\overleftrightarrow{G}_{\hat{\mathbf{k}}}  = \left( \begin{array}{ccc} 0 & -k_z & k_y \\ k_z & 0 & -k_x \\ -k_y & k_x & 0 \end{array} \right), \qquad \lambda_{\hat{\mathbf{k}} \alpha} = \frac{\mu_{\mathbf{k} \alpha}}{ik}
\end{equation}
and $\mathbf{k} = k \, (k_x,k_y,k_z) = k \, \hat{\mathbf{k}}$ with $k_x^2 + k_y^2 + k_z^2 = 1$.  The eigenvalues and eigenvectors are:
\begin{equation}
\begin{split}
\mathbf{a}_{\hat{\mathbf{k}} 0} &= \hat{\mathbf{k}}, \\
\lambda_{\hat{\mathbf{k}} 0} &= 0, \\
\mu_{\mathbf{k} 0} &= 0, \\
\mathbf{a}_{\hat{\mathbf{k}} \pm} &= \frac{1}{\sqrt{2(1-k_x^2)}} \left( \begin{array}{c} k_x^2-1 \\ k_x k_y \mp i k_z \\ k_x k_z \pm i k_y \end{array} \right), \\
\lambda_{\hat{\mathbf{k}} \pm} &= \mp i \\
\text{and} \quad \mu_{\mathbf{k} \pm} &= \pm k.
\end{split}
\end{equation}
The following Hodge decomposition for $\mathbf{v}$ on $B^3$ can now be obtained
\begin{equation}
\mathbf{v}(\mathbf{q}) = \mathbf{C}_0 + \sum_{\begin{array}{c} \mathbf{n} \ne \vec{0} \\ n_x,n_y,n_z \ge 0 \end{array}} \; \sum_{\alpha=0,\pm} {C_{\mathbf{n} \alpha} \; \vec{\xi}_{\mathbf{n} \alpha}(\mathbf{q},\beta_{\mathbf{n} \alpha})},
\end{equation}
where $C_{\mathbf{n} \alpha}$ and $\beta_{\mathbf{n} \alpha}$ are real constants, $\mathbf{C}_0$ is a constant vector,
\begin{equation}
\vec{\xi}_{\mathbf{n} 0}(\mathbf{q},\beta_{\mathbf{n} 0}) = \sqrt{2}\; \mathbf{a}_{\hat{\mathbf{k}} 0} \cos(\mathbf{q} \cdot  \mathbf{k} + \beta_{\mathbf{n} 0})
\end{equation}
and
\begin{equation}
\vec{\xi}_{\mathbf{n} \pm}(\mathbf{q},\beta_{\mathbf{n} \pm}) = \sqrt{2} \; \left[ \text{Re} \left( \mathbf{a}_{\hat{\mathbf{k}} \pm} \right) \cos(\mathbf{q} \cdot  \mathbf{k} + \beta_{\mathbf{n} \pm}) - \text{Im} \left( \mathbf{a}_{\hat{\mathbf{k}} \pm} \right) \sin(\mathbf{q} \cdot  \mathbf{k} + \beta_{\mathbf{n} \pm}) \right].
\end{equation}
It is easy to verify that the vectors $\vec{\xi}_{\mathbf{n} \alpha}$ are orthogonal,
\begin{equation}
\left( \vec{\xi}_{\mathbf{n} \alpha}, \vec{\xi}_{\mathbf{n}' \alpha'} \right) = \delta_{\mathbf{n},\mathbf{n}'} \; \delta_{\alpha,\alpha'},
\end{equation}
with respect to the inner product
\begin{equation}
\left(\mathbf{a},\mathbf{b} \right) = \frac{1}{L_x L_y L_z}  \int{\mathbf{a} \cdot \mathbf{b} \; \; d^3q}.
\end{equation}
In addition, the vector calculus properties of these vectors can be summarized as follows:
\begin{equation}
\begin{split}
\vec{\nabla} \cdot \mathbf{C}_0 &= 0, \\
\vec{\nabla} \times \mathbf{C}_0 &= \vec{0}, \\
\vec{\nabla} \cdot \vec{\xi}_{\mathbf{n} 0} &= -k \sqrt{2} \sin(\mathbf{q} \cdot  \mathbf{k} + \beta_{\mathbf{n} 0}), \\
\vec{\nabla} \times \vec{\xi}_{\mathbf{n} 0} &= \vec{0}, \\
\vec{\nabla} \cdot \vec{\xi}_{\mathbf{n} \pm} &= 0 \\
\text{and} \qquad \vec{\nabla} \times \vec{\xi}_{\mathbf{n} \pm} &= \pm k  \vec{\xi}_{\mathbf{n} \pm}.
\end{split}
\end{equation}
The harmonic, irrotational and the solenoidal components of $\mathbf{v}$ are respectively:
\begin{equation}
\begin{split}
\mathbf{v}_h &= \mathbf{C}_0, \\
\mathbf{v}_i &=  \sum_{\begin{array}{c} \mathbf{n} \ne \vec{0} \\ n_x,n_y,n_z \ge 0 \end{array}} {C_{\mathbf{n} 0} \; \vec{\xi}_{\mathbf{n} 0}(\mathbf{q},\beta_{\mathbf{n} 0})} \\
\text{and} \qquad \mathbf{v}_s &=  \sum_{\begin{array}{c} \mathbf{n} \ne \vec{0} \\ n_x,n_y,n_z \ge 0 \end{array}} {C_{\mathbf{n}+} \; \vec{\xi}_{\mathbf{n}+}(\mathbf{q},\beta_{\mathbf{n}+}) + C_{\mathbf{n}-} \; \vec{\xi}_{\mathbf{n}-}(\mathbf{q},\beta_{\mathbf{n}-})}.
\end{split}
\end{equation}
If $\mathbf{v}$ is restricted to the solenoidal vectors, the minimum eigenvalue of curl is $m=2 \pi / L_\text{max}$, where $L_\text{max}= \max(L_x,L_y,L_z)$.  Under any Hamiltonian flow on $B^3$, the enstrophy has a lower bound so that
\begin{equation}
\mathscr{N} \ge \frac{2 \pi}{L_\text{max}} |\mathscr{H}|.
\end{equation}
Written in terms of the pullback of Sec.~\ref{sec:pullback}, this condition is
\begin{equation}
\int {\left( \vec{\nabla} \times \mathbf{p} \right)^2  \; d^3q} \ge \frac{2 \pi}{L_\text{max}} \left| \int {\mathbf{p} \cdot \left( \vec{\nabla} \times \mathbf{p} \right)  \; d^3q} \right|.
\end{equation}
The solenoidal energy will have an upper bound so that
\begin{equation}
\mathscr{T}_0 \le \frac{L_\text{max}}{2 \pi} |\mathscr{H}|.
\end{equation}
The pullback of this condition is
\begin{equation}
\int {\mathbf{p}_s^2  \; d^3q} \le \frac{L_\text{max}}{2 \pi} \left| \int {\mathbf{p} \cdot \left( \vec{\nabla} \times \mathbf{p} \right)  \; d^3q} \right|.
\end{equation}

Now let us turn our attention to a compact manifold with boundary.  Consider a cylinder of radius $a$ and length $L$.  Because $\partial V \ne 0$, one must impose a boundary condition if helicity is to be conserved and the surface terms in the variational calculation are to vanish.  A simple way to do this is to require $\Omega^2|_{\partial V} = 0$ initially.  Since $\mathscr{L}_\mathbf{u} \Omega^2 = 0$, $\Omega^2|_{\partial V} = 0$ for all time.  Under this condition,
\begin{equation}
\frac{d \mathscr{H}}{dt} = \int_{\partial V} {\left( i_\mathbf{u}\Lambda - H \right) \wedge \Omega} = 0.
\end{equation}
Also, the surface term in the variational calculation is
\begin{equation}
\int_{\partial V} {\delta \Lambda \wedge \left(2*\Omega - \mu \Lambda \right)} = \int_{\partial V} {\mu \, \Lambda \wedge \delta \Lambda}.
\end{equation}
The variation is constrained so that $\delta \Omega|_{\partial V} = d\left( \delta \Lambda|_{\partial V} \right)= 0$ (\textit{i.e.,} $\delta \Lambda|_{\partial V}$ is closed).  We assume that an additional condition on the topology of $\partial V$ is met.  The first Betti number of $\partial V$ is zero.  This is equivalent to all one chains on $\partial V$ being homologous to zero.  The cylinder obviously meets this condition.  The variation $\delta \Lambda|_{\partial V}$ is therefore exact and can be written as $\delta \Lambda|_{\partial V} = d\alpha^0$.  We can now further reduce the surface term to
\begin{equation}
\int_{\partial V} {\mu \, \Lambda \wedge \delta \Lambda} = \mu \int_{\partial V} {\Lambda \wedge d\alpha^0} = \mu \int_{\partial V} { \alpha^0\Omega - d(\alpha^0\Lambda)} = - \mu \int_{\partial(\partial V)} {\alpha^0 \Lambda} = 0.
\end{equation}

Since helicity is conserved, we now need to solve the eigenvalue equation for curl on this manifold under the above boundary condition.  A simple way to estimate the bound on the eigenvalue is to apply the result for $B^3$.  Imbed the cylinder in a box with sides of length $L_x = L_y = 2a$ and $L_z = L$, and close it into $T^3$ topology by applying periodic boundary conditions.  We now find that the minimum solenoidal eigenvalue for the cylinder is more than $m_0 = 2\pi / \max(2a,L)$.

To find the minimum solenoidal eigenvalue and the corresponding eigenvector, we need to obtain solutions to
\begin{equation}
\vec{\nabla} \times \mathbf{a} = \mu \mathbf{a}
\end{equation}
in cylindrical coordinates such that $\mathbf{a}$ is solenoidal (\textit{i.e.,} $\mathbf{a} = \vec{\nabla} \times \mathbf{b}$ for some $\mathbf{b}$).  These have been previously been found \citep{manheimer1984mhd,chandrasekhar1957force,montgomery1978three} to be
\begin{equation}
\begin{split}
a^{mk}_r &= \frac{-1}{\sqrt{\mu^2 - k^2}}\left[ k{J'}_m(y) + \frac{\mu m}{y} J_m(y) \right] \sin(m\theta + kz), \\
a^{mk}_\theta &= \frac{-1}{\sqrt{\mu^2 - k^2}}\left[ \mu {J'}_m(y) + \frac{mk}{y} J_m(y) \right] \cos(m\theta + kz) \\
\text{and} \qquad a^{mk}_z &= J_m(y) \cos(m\theta + kz),
\end{split}
\end{equation}
where $J_m(z)$ is the Bessel function and $y = r \sqrt{\mu^2 - k^2}$. 

Applying the boundary condition $\mathbf{a}|_{\partial V} = \vec{0}$, restricts one to solutions such that $m=0$, $kL=n\pi$, $n \ne0$ and $J_1(a \sqrt{\mu^2 - k^2}) = 0$.  This gives
\begin{equation}
\mu_{pn} = \pm \sqrt{\left( \frac{y_p}{a} \right)^2 + \left( \frac{n\pi}{L} \right)^2},
\end{equation}
where $J_1(y_p)=0$.  The minimum solenoidal eigenvalue is
\begin{equation}
m = |\mu_{11}| = \sqrt{\left( \frac{y_1}{a} \right)^2 + \left( \frac{\pi}{L} \right)^2},
\end{equation}
which can be shown to be greater than the bound $m_0 = 2\pi / \max(2a,L)$ found earlier.  As expected $m>m_0$.  This constant $m$ sets the lower bound on enstrophy and the upper bound on solenoidal energy.

\section{Physical examples}\label{sec:phys_example}
We discuss three examples of Hamiltonian systems, each of increasing complexity.

\subsection{Continuous charged dust}\label{sec:charged_dust}
\begin{figure}[ht]
\center\includegraphics[width=20pc]{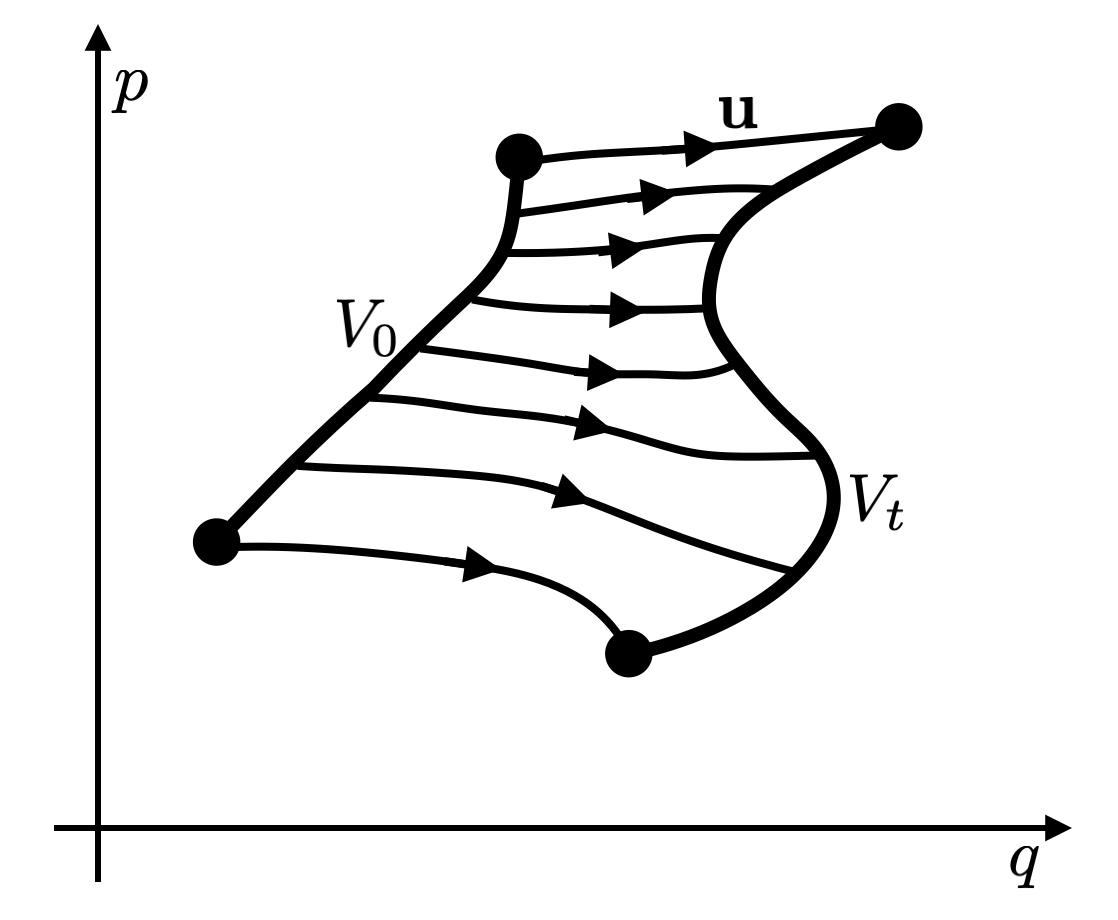}
\caption{\label{fig:flow}Flow of a graph $V$ on phase space $T^*M$.}
\end{figure}

The first is a system of ``continuous charged dust''.  This system consists of many small dust particles of mass and charge $+1$.  The state of the system is given by the density of the dust $n(q,t)$ and its velocity $\mathbf{v}(q,t)$ as functions of position $q=(x,y,z)$ and time $t$.  For simplicity, we write the state at time $t$ as
\begin{equation}
s(t) = ( \mathbf{v}(q,t),n(q,t)).
\end{equation}
We will consider only electrostatic interaction so that we can write the Lagrange function as
\begin{equation}
L[s](q,\mathbf{v},t) = \frac{v^2}{2} - V_0[s](q) - V_\text{ext}(q,t) - V_c(q),
\end{equation}
where
\begin{equation}
V_0[s](q) = \int {\frac{n(q',t)}{|q-q'|} dq'}
\end{equation}
is the potential due to the other dust particles, $V_\text{ext}$ is an external potential applied to the system,
\begin{equation}
V_c(q) = \left\{ \begin{array}{lc} 0 & |q| \le q_0 \\ \infty & |q| > q_0 \end{array} \right.
\end{equation}
is a potential to contain the system of charges, and
\begin{equation}
\mathbf{v} =  \frac{dx}{dt} \frac{\partial}{\partial x} + \frac{dy}{dt} \frac{\partial}{\partial y}+ \frac{dz}{dt} \frac{\partial}{\partial z} = v^x \frac{\partial}{\partial x} + v^y \frac{\partial}{\partial y}+ v^z \frac{\partial}{\partial z}
\end{equation}
is the tangent vector to $M$.  Both $V_0$ and $L$ are functionals of the current state of the system.  We now use a Legendre transformation to obtain the Hamiltonian
\begin{equation}
\label{eqn:ham_dust}
H[s](q,p,t) = \frac{p^2}{2} + V_0[s](q) + V_\text{ext}(q,t) + V_c(q),
\end{equation}
where $p=\partial L / \partial \mathbf{v} = (p_x,p_y,p_z) = (v_x,v_y,v_z)$.  The evolution of the system is determined by Hamilton's equations
\begin{equation}
\label{eqn:ham_eqn_dust}
\begin{split}
\frac{dp}{dt} &= - \frac{\partial H}{\partial q}, \\
\frac{dq}{dt} &= \frac{\partial H}{\partial p}, \\
\frac{dH}{dt} &=  \frac{\partial H}{\partial t} 
\end{split}
\end{equation}
and mass conservation
\begin{equation}
\frac{dn}{dt} = - n \vec{\nabla} \cdot  \mathbf{v} .
\end{equation}
Given the initial state of the system, $s_0$, we can solve for $s(s_0,t)$.  We can then substitute into Eq.~\ref{eqn:ham_dust} to get $H[s_0](q,p,t)$.  Now consider extended phase space where $Q=(t,x,y,z)$ and $P=(-H,p_x,p_y,p_z)=(-H,v_x,v_y,v_z)$.  The new Hamiltonian is $H'(Q,P)=0$, and the Poincar\'e 1-form is $\Lambda^1 = p \, dq -H \, dt$.  Equations~\ref{eqn:ham_eqn_dust} can now be rewritten in the familiar form $i_\mathbf{u} \Omega = 0$.  The flow of a graph $V$ is then generated by the Hamiltonian vector field $\mathbf{u}$ (see Fig.~\ref{fig:flow}).  A point to note is that $\mathbf{u}$ is a functional of $s_0$, but no matter what $s_0$ one picks, the flow is still Hamiltonian.

The implications of this flow in terms of familiar vector expressions are now examined.  The pullback of the Poincar\'e 2-form and the helicity 3-form are
\begin{equation}
\beta^* \Omega^2 = \omega^2_{\vec{\nabla} \times \mathbf{v}} + \left( \omega^1_{-\partial_t \mathbf{v} - \vec{\nabla} H} \right) \wedge dt
\end{equation}
and
\begin{equation}
\beta^* K^3 = \left( \mathbf{v} \cdot \vec{\nabla} \times \mathbf{v} \right) \text{vol}^3 - \left( \omega^2_{H \vec{\nabla} \times \mathbf{v} + \mathbf{v} \times \vec{\nabla} H + \mathbf{v} \times \partial_t \mathbf{v}} \right) \wedge dt.
\end{equation}
For a graph $V$ with a constant time coordinate;  one can write helicity, enstrophy and solenoidal energy as
\begin{equation}
\begin{split}
\mathscr{H} &= \int {\mathbf{v} \cdot \left(\vec{\nabla} \times \mathbf{v} \right) \; d^3q}, \\
\mathscr{N} &= \int {\left(\vec{\nabla} \times \mathbf{v} \right)^2 \; d^3q} \\
\text{and} \qquad \mathscr{T}_0 &=  \int {\mathbf{v}^2_s  \; d^3q},
\end{split}
\end{equation}
respectively.  If $\partial V = 0$ or $L (\hat{\mathbf{n}} \cdot \vec{\nabla} \times \mathbf{v} )|_{\partial V} = 0$, where $\hat{\mathbf{n}}$ is a unit normal to $V$, we know from Sec.~\ref{sec:defs} that $\mathscr{H}$ is conserved as the graph undergoes the Hamiltonian flow generated by $\mathbf{u}$.  Furthermore, since the dust is constrained to flow within the box of size $q_0$, we know that enstrophy has a lower bound of order $|\mathscr{H}| / q_0$ and solenoidal energy has an upper bound of order $|\mathscr{H}| q_0$.  These bounds are independent of the choice of the external time dependent potential $V_\text{ext}$.

\subsection{Perfect fluid}\label{sec:perfect_fluid}
A more familiar continuous system that can be written in terms of a Hamiltonian flow is a perfect fluid \citep{landau1987fluid}.  The state of the system is given by the fluid velocity $\mathbf{v}(q,t)$, density $n(q,t)$ and pressure $P(q,t)$ as functions of position $q=(x,y,z)$ and time $t$, that is,
\begin{equation}
s(t) = (P(q,t),n(q,t),\mathbf{v}(q,t)).
\end{equation}
The equations which govern the evolution of a perfect fluid are Euler's Equation
\begin{equation}
\label{eqn:eulers}
\frac{d \mathbf{v}}{dt} = - \frac{1}{n} \vec{\nabla} P,
\end{equation}
the equation of mass conservation,
\begin{equation}
\label{eqn:mass_con}
\frac{\partial n}{\partial t} + \vec{\nabla} \cdot \left( n \mathbf{v} \right) = 0,
\end{equation}
and an equation of state such as:
\begin{equation}
\label{eqn:eos}
\begin{split}
\vec{\nabla} \cdot \mathbf{v} = 0 \qquad  &  \text{(i.e., incompressible),} \\
\frac{d}{dt} \left( P n^{-\gamma} \right) = 0  \qquad &  \text{(i.e., adiabatic)} \\
\text{or}  \qquad  \frac{d}{dt} \left( \frac{P}{n} \right) = 0  \qquad &  \text{(i.e., isothermal).}
\end{split}
\end{equation}
In order to write Euler's equation as Hamilton's equations, we introduce the enthalpy per particle
\begin{equation}
\mu \equiv \frac{1}{n} \left( P+e \right),
\end{equation}
where $e$ is the energy density.  A thermodynamic identity $\mu = de / dn$ shows us that $e$ and $\mu$ are not independent quantities.  This allows us to write $\mu$ as a functional of $P$ and $n$ with an explicit dependence on $q$, that is,
\begin{equation}
\mu = \mu[s](q).
\end{equation}
Now consider the Lagrange function
\begin{equation}
L_\text{F}[s](q,\mathbf{v}) = \frac{v^2}{2} - \mu[s](q).
\end{equation}
A Legendre transformation yields the Hamiltonian
\begin{equation}
\label{eqn:ham_fluid}
H_\text{F}[s](q,p) = \frac{p^2}{2} + \mu[s](q)
\end{equation}
with $p=\partial L_\text{F} / \partial \mathbf{v} = (p_x,p_y,p_z) = (v_x,v_y,v_z)$.  To eliminate the functional dependence on $s$, solve Eqs.~\ref{eqn:eulers},~\ref{eqn:mass_con} and~\ref{eqn:eos} for $s=s(s_0,t)$.  Now substitute into Eq.~\ref{eqn:ham_fluid} to yield
\begin{equation}
H_\text{F}[s_0](q,p,t) = \frac{p^2}{2} + \mu[s(s_0,t)](q).
\end{equation}
The equations
\begin{equation}
\begin{split}
\frac{dp}{dt} &= -\frac{\partial H_\text{F}}{\partial q} \\
\text{and} \qquad \frac{dq}{dt} &= \frac{\partial H_\text{F}}{\partial p}
\end{split}
\end{equation}
can be shown to be equivalent to Euler's equation.  Because of the explicit time dependence of $H_\text{F}$, we must examine the flow of the graph on extended phase space.  This evolution is governed by $i_\mathbf{u} \Omega^2_\text{F} = 0$ where $\Lambda^1_\text{F} = p \; dq - H_\text{F} \; dt$.  The rest of the analysis is identical to that for the ``continuous charged dust'' if the fluid is constrained to move inside a box of size $q_0$.

\subsection{MHD}\label{sec:mhd}
The last example of a system with a conserved helicity is MHD \citep{krall1986principles}.  The state of the system is given by the density $n(q,t)$, pressure $P(q,t)$, fluid velocity $\mathbf{v}(q,t)$, charge density $n_q(q,t)$, current $\mathbf{j}(q,t)$, scalar potential $\varphi(q,t)$ and vector potential $\mathbf{A}(q,t)$ as functions of position and time, that is,
\begin{equation}
s(t) = ( \; n(q,t),P(q,t),\mathbf{v}(q,t), n_q(q,t), \mathbf{j}(q,t), \varphi(q,t), \mathbf{A}(q,t) \; ).
\end{equation}
For convenience in writing the MHD equations; we set $c=1$, define the magnetic field by
\begin{equation}
\mathbf{B} \equiv \vec{\nabla} \times \mathbf{A}
\end{equation}
and define the electric field by
\begin{equation}
\mathbf{E} \equiv - \frac{\partial \mathbf{A}}{\partial t} - \vec{\nabla} \varphi.
\end{equation}

Two MHD equations which govern the evolution of the system we will examine in further detail.  The first is Ohm's law
\begin{equation}
\mathbf{E} + \mathbf{v} \times \mathbf{B} = \frac{\mathbf{j}}{\sigma}  \approx 0
\end{equation}
where we have assumed infinite conductivity $\sigma$.  The second is the equation of force balance
\begin{equation}
\label{eqn:fbal}
n \frac{d \mathbf{v}}{dt} = \mathbf{j} \times \mathbf{B} + n_q \mathbf{E} - \vec{\nabla} P.
\end{equation}
We also have the equation of mass conservation
\begin{equation}
\frac{\partial n}{\partial t} + \vec{\nabla} \cdot (n \mathbf{v}) = 0,
\end{equation}
the equation of charge conservation
\begin{equation}
\frac{\partial n_q}{\partial t} + \vec{\nabla} \cdot \mathbf{j} = 0,
\end{equation}
and Maxwell's equations
\begin{equation}
\begin{split}
\vec{\nabla} \cdot \mathbf{E} &= 4\pi n_q \\
\text{and} \qquad \vec{\nabla} \times \mathbf{B} &= 4\pi \mathbf{j}.
\end{split}
\end{equation}
The system of equations is completed by an equation of state such as those which appear in Eq.~\ref{eqn:eos}.  As with the previous two examples we can solve the system of equations to obtain $s(s_0,t)$.

What we now wish to do is rewrite Ohm's law and the force balance equations in terms of interior products of Poincar\'e 2-forms.  The Hamiltonian structure of Ohm's law can be uncovered by considering the Lagrangian function
\begin{equation}
L_\text{em}[s_0](Q,\mathbf{U}) = A_a U^a,
\end{equation}
where $A=(-\varphi,A_x,A_y,A_z)$ is the covector form of the electromagnetic 4-potential and $Q=(t,x,y,z)$ is a point in Minkowski space with the metric
\begin{equation}
((g_{ab})) = \left( \begin{array}{cccc} -1 & 0 & 0 & 0 \\ 0 & 1 & 0 & 0 \\ 0 & 0 & 1 & 0 \\ 0 & 0 & 0 & 1 \end{array} \right),
\end{equation}
and
\begin{equation}
\mathbf{U} = \mathbf{v} + \frac{\partial}{\partial t}
\end{equation}
is the tangent vector to $M$.  The canonical momentum is $P = \partial L_{em} / \partial \mathbf{U} = A$ and the Hamilton function is 
\begin{equation}
H_\text{em}[s_0](Q,P) = 0.
\end{equation}
The Poincar\'e one form is written as $\Lambda^1_\text{em} = A_a \, dQ^a = \mathbf{A} \cdot d\mathbf{q} - \varphi \, dt$.  We can now write Hamiltons's equations as
\begin{equation}
\label{eqn:ham_em}
i_\mathbf{u} \Omega^2_\text{em} = 0.
\end{equation}
This can be shown to be equivalent to Ohm's law in the following way.  Consider the pullback $\beta^*$ from $\Lambda^p(T^*M)$ to $\Lambda^p(M)$.  Apply it to Eq.~\ref{eqn:ham_em} to yield
\begin{equation}
\label{eqn:ham_em_pb}
0 = i_\mathbf{U} \left( \beta^* \Omega^2_\text{em} \right),
\end{equation}
where
\begin{equation}
\beta^* \Omega^2_\text{em} = \omega^2_\mathbf{B} + \omega^1_\mathbf{E} \wedge dt
\end{equation}
is the electromagnetic field 2-form.  Further simplification of Eq.~\ref{eqn:ham_em_pb} yields
\begin{equation}
\omega^1_{\mathbf{E} + \mathbf{v} \times \mathbf{B}} - \mathbf{v} \cdot \left( \mathbf{E} + \mathbf{v} \times \mathbf{B} \right) dt = 0.
\end{equation}
This equation, when written in vector form, is just Ohm's law.  Since we now have expressed Ohm's law in Hamiltonian form, we can say that magnetic helicity will be conserved if $\partial V = 0$ or $L_\text{em} \Omega^2_\text{em}|_{\partial V} = 0$.  To relate this conservation to more familiar expressions, we pullback the magnetic helicity 3-form.  The result is
\begin{equation}
\beta^* K^3 = ( \mathbf{A} \cdot \mathbf{B} ) \text{vol}^3 - \left( \omega^2_{\varphi \mathbf{B} + \mathbf{E} \times \mathbf{A}} \right) \wedge dt.
\end{equation}
If the graph $V$ has constant time;  then magnetic helicity, enstrophy and solenoidal energy are:
\begin{equation}
\begin{split}
\mathscr{H} &= \int {\mathbf{A} \cdot \mathbf{B} \; \; d^3q}, \\
\mathscr{N} &= \int {\mathbf{B}^2 \; \; d^3q} \\
\text{and} \qquad  \mathscr{T}_0 &= \int {\mathbf{A}^2_s \; \; d^3q}.
\end{split}
\end{equation}
The boundary condition for such a $V$ required for magnetic helicity conservation is $\partial V = 0$ or $L_\text{em} (\mathbf{B} \cdot \hat{\mathbf{n}} )|_{\partial V} = 0$.

We now turn our attention to the force balance equation.  While we are not able to write this in terms of Hamilton's equations, we are able to write it in a form sufficiently close to Hamiltonian so that a quantity called cross helicity will be conserved if certain boundary conditions are met.  First notice the similarity between the force balance equation and Euler's equation.  The only difference is the $\mathbf{j} \times \mathbf{B} + n_q \mathbf{E}$ term in the force balance equation.  Inspired by this similarity, we use the same Poincar\'e 2-form we used for the fluid $\Omega^2_\text{F}$ to rewrite the force balance equation as
\begin{equation}
\label{eqn:fbal_em}
i_\mathbf{U} \Omega^2_\text{F} = - i_\mathbf{J} \Omega^2_\text{em},
\end{equation}
where
\begin{equation}
\mathbf{J} = \frac{\mathbf{j}}{n} + \frac{n_q}{n} \frac{\partial}{\partial t} = \frac{n_q}{n} \frac{\partial}{\partial t} + \frac{j^x}{n} \frac{\partial}{\partial x} + \frac{j^y}{n} \frac{\partial}{\partial y} + \frac{j^z}{n} \frac{\partial}{\partial z}.
\end{equation}
We adopt the convention that $\beta^*_\text{em}$ operates on $\Lambda^1_\text{em}$ and $\beta^*_\text{F}$ operates on $\Lambda^1_\text{F}$ whenever they appear.  This is necessary since $\Lambda^1_\text{em}$ and $\Lambda^1_\text{F}$ are forms acting on different cotangent bundles which have the same base manifold $M$.  Therefore, to have expressions such as $\Lambda^1_\text{em} \wedge \Lambda^1_\text{F}$ make sense, both forms must be pulled-back to the same base manifold.  The equivalence of Eq.~\ref{eqn:fbal_em} and the force balance equation, Eq.~\ref{eqn:fbal}, can be seen by reducing Eq.~\ref{eqn:fbal_em} to
\begin{equation}
\label{eqn:fbal_simple}
\omega^1_\mathbf{d} - (\mathbf{v} \cdot \mathbf{d} ) \; dt = 0,
\end{equation}
where
\begin{equation}
\mathbf{d} \equiv \frac{d \mathbf{v}}{dt} + \frac{1}{n} \left( \vec{\nabla} P - \mathbf{j} \times \mathbf{B} - n_q \mathbf{E} \right).
\end{equation}
To express the second term in Eq.~\ref{eqn:fbal_simple} as it appears, we have used the fact that Ohm's law is satisfied.

The 1-form $i_\mathbf{J} \Omega^2_\text{em}$ need not be exact so that Eq.~\ref{eqn:fbal_em} is not in the form of Hamilton's equations.  Because of the non-exact part of $i_\mathbf{J} \Omega^2_\text{em}$, fluid helicity $\mathscr{H}_\text{F} = \int_V \Lambda^1_\text{F} \wedge \Omega^2_\text{F}$ will not be conserved.  Despite this, we can define the cross helicity
\begin{equation}
\mathscr{H}_c \equiv \int_V {\Lambda^1_\text{F} \wedge \Omega^2_\text{em}},
\end{equation}
which is conserved if certain boundary conditions are met.  Here, $V$ is a volume in Minkowski space.  The reason for this conservation can be seen by taking the time derivative of $\mathscr{H}_c$ as follows:
\begin{equation}
\label{eqn:cross_helicity}
\begin{split}
\frac{d \mathscr{H}_c}{dt} &= \int_V {\mathscr{L}_\mathbf{U} \left( \Lambda^1_\text{F} \wedge \Omega^2_\text{em} \right)} \\
&= \int_V {\left( \mathscr{L}_\mathbf{U} \Lambda^1_\text{F} \right) \wedge \Omega^2_\text{em} + \Lambda^1_\text{F} \wedge \left( \mathscr{L}_\mathbf{U} \Omega^2_\text{em}\right) } \\
&= \int_V {\left( \mathscr{L}_\mathbf{U} \Lambda^1_\text{F} \right) \wedge \Omega^2_\text{em} }.
\end{split}
\end{equation}
The second term with $ \mathscr{L}_\mathbf{U} \Omega^2_\text{em}$ is zero since Ohm's law is Hamiltonian.  We now apply Cartan's formula and Stoke's theorem to Eq.~\ref{eqn:cross_helicity} and express the time derivative as
\begin{equation}
\label{eqn:cross_helicity_dt}
\frac{d \mathscr{H}_c}{dt} = \int_{\partial V} {\left( i_\mathbf{U} \Lambda^1_\text{F} \right) \Omega^2_\text{em}} + \int_V {\left( i_\mathbf{U} \Omega^2_\text{F} \right) \wedge \Omega^2_\text{em}}.
\end{equation}
For the case of a Hamiltonian system, we would substitute $-dH$ for $ i_\mathbf{U} \Omega^2_\text{F} $.  We can not do this because of the non-exact $i_\mathbf{J} \Omega^2_\text{em}$ term in the force balance equation.  What we can do is substitute $-i_\mathbf{J} \Omega^2_\text{em}$ for $i_\mathbf{U} \Omega^2_\text{F}$ in the second term of Eq.~\ref{eqn:cross_helicity_dt}. By use of Ohm's law, one can now show that this term equals zero.  This leaves us with
\begin{equation}
\frac{d \mathscr{H}_c}{dt} = \int_{\partial V} { L_\text{F} \, \Omega^2_\text{em}},
\end{equation}
where $ L_\text{F} \equiv i_\mathbf{U} \Lambda^1_\text{F} $. Therefore, if either $\partial V = 0$ or $L_\text{F} \, \Omega^2_\text{em}|_{\partial V} = 0$, cross helicity will be conserved.  The pullback of the cross helicity 3-form is
\begin{equation}
\beta^* K^3_c = \left(\mathbf{v} \cdot \mathbf{B} \right) \text{vol}^3 + \left( \omega^2_{-H_\text{F} \mathbf{B} + \mathbf{v} \times \mathbf{E}} \right) \wedge dt.
\end{equation}
For a graph $V$ with constant time, the cross helicity is
\begin{equation}
\mathscr{H}_c = \int_V \mathbf{v} \cdot \mathbf{B} \; \; d^3q.
\end{equation}
The boundary condition so that cross helicity be conserved on $V$ is $\partial V = 0$ or $L_\text{F} ( \mathbf{B} \cdot \hat{\mathbf{n}} )|_{\partial V} = 0$ where $\hat{\mathbf{n}}$ is a unit normal to $\partial V$.

\section{Conclusions}\label{sec:conclusions}
We have defined helicity density $K^3$ as the natural 3-form on $T^*M$.  Under Hamiltonian flow and certain boundary conditions, helicity,
\begin{displaymath}
\mathscr{H} \equiv \int_V K^3,
\end{displaymath}
is conserved.  This limits the class of configuration obtainable through Hamiltonian flow.  The limited class of configurations has a lower bound on enstrophy and an upper bound on solenoidal energy.  These bounds are set by the minimum solenoidal eigenvalue of the curl operator, $d*$, on the three dimensional manifold.  If helicity were not conserved, these bounds would not exist.

\begin{acknowledgments}
We would like to thank M.H Freedman and T.T. Frankel for many enlightening discussions.  The work was supported by a National Science Foundation Graduate Fellowship and National Science Foundation Grant No. PHY87-06358.  This report was prepared by Sandia National Laboratories, a multimission laboratory managed and operated by National Technology and Engineering Solutions of Sandia LLC, a wholly owned subsidiary of Honeywell International Inc. for the U.S. Department of Energy's National Nuclear Security Administration under contract DE-NA0003525.  This work was conducted while M.E.G. was a graduate student at University of California, San Diego.  The technical content of this report is contained in an unpublished manuscript completed during September 1990.  This is a Sandia National Laboratories Technical Report, SAND2019-14731.
\end{acknowledgments}

\bibliography{helicity_SAND_References.bib}

\end{document}